# Quantum Spectroscopy with Undetected Photons for Biomolecular Sensing in the Mid-Infrared

Mahya Mohammadi[1,2*], Meryem-Nur Duman[1,2], Isa Ahmadalidokht[1,2], Mohammad Sadraeian[2,3], Christopher G. Poulton[1], Alexander S. Solntsev[1,2], Irina V. Kabakova[1,2*]

[1.] School of Mathematical and Physical Sciences (MaPS), Faculty of Science, University of Technology Sydney, Sydney, NSW 2007, Australia

[2.] ARC Centre of Excellence in Quantum Biotechnology (QUBIC), University of Technology Sydney, Sydney, NSW 2007, Australia

[3] Institute for Biomedical Materials and Devices (IBMD), Faculty of Science, University of Technology Sydney, Sydney, NSW 2007, Australia

*Correspondence to: Mahya.Mohammadi@student.uts.edu.au, Irina.Kabakova@uts.edu.au

## Abstract

We investigate quantum spectroscopy with undetected photons for protein detection in the mid-infrared spectral region. Classical Fourier-transform infrared spectroscopy of protein samples (bovine serum albumin and N-terminal pro-brain natriuretic peptide) is used as reference to define the sample's mid-infrared absorption, which is then embedded in a numerical model of a double-pass quantum interferometer. We analyse parameters that influence visibility of the interference pattern formed by the signal beams, including the length of nonlinear crystal, sample length and mirror-sample distance. This leads us to a practical quantum spectrometer design with optimal image contrast at the specific amide I-II spectral bands. The simulated visibility spectra reproduce nearly identically the protein absorption features in the mid-IR and reveal temperature-induced changes to the protein secondary structure. Overall, this provides practical design rules for future quantum bio-spectroscopy applications that use only visible wavelength sources and detectors.

## 1. Introduction

The mid-infrared (MIR) range of the electromagnetic spectrum, corresponding to wavelengths between approximately 2.5 and 25 μm (4000 to 400 cm$^{-1}$), is of great importance for environmental sensing, biomolecular detection, and defence applications [1, 2]. This spectral region encompasses the fundamental vibrational modes of most chemical bonds (including C–H, O–H, and C=O stretching vibrations) which give rise to characteristic molecular absorption bands. Precise measurement of these absorption bands with high sensitivity and spectral resolution enables direct characterisation of molecular fingerprints in unknown substances. For proteins, the MIR operation offers an opportunity to study changes in secondary structure in response to environmental factors such as temperature, pH, and ion concentration [3, 4].

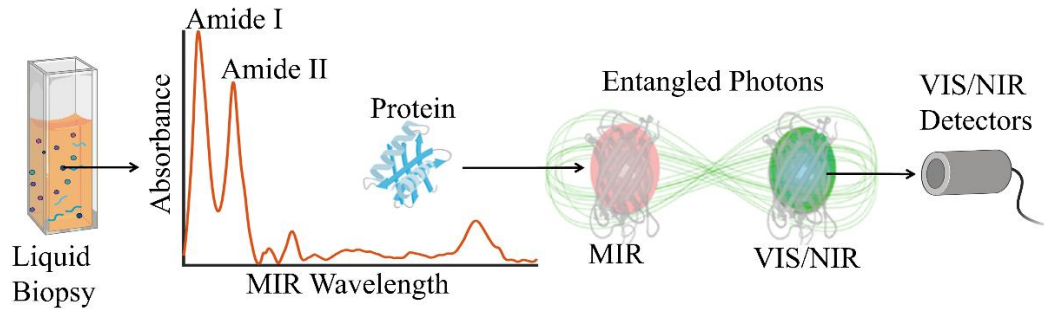

**Figure 1.** Conceptual scheme for retrieving mid-infrared (MIR) spectroscopic information of proteins using visible (VIS) / near-infrared (NIR) photons. A liquid biopsy sample containing proteins exhibits characteristic vibrational absorption features in the MIR spectral region, such as the Amide I and Amide II bands. In the proposed approach, correlated photon pairs are generated through spontaneous parametric down-conversion (SPDC), producing an idler photon in the MIR range and a signal photon in the VIS/NIR range. The MIR idler photon interacts with the protein sample and acquires information about its absorption. Through quantum correlations between the photon pairs, this interaction modifies the interference properties of the detected signal photons. As a result, the MIR spectroscopic response of the sample can be retrieved by measuring the VIS/NIR photons with conventional detectors, enabling indirect access to molecular vibrational information without requiring direct MIR detection.

Several standard methods exist to analyse molecular absorption in the MIR, with Fourier Transform Infrared (FTIR) spectroscopy being the most widely used and considered the industry standard for label-free MIR absorption measurements. FTIR spectroscopy is particularly powerful for studying protein structure and monitoring conformational changes [3, 4]. Proteins exhibit nine characteristic IR absorption bands, known as amide bands (A, B, and I–VII). Among these, the amide I band (~5.88–6.25 μm) has the strongest absorption and is highly sensitive to secondary structural components such as α-helices, β-sheets, and β-turns. Subtle shifts in vibrational frequencies arise from differences in hydrogen bonding and amide bond geometry among these structures [5]. However, the applicability of FTIR is constrained by factors such as strong water absorption in the MIR region, which reduces sensitivity for dilute samples [3]. Other MIR methods, based on tuneable MIR lasers and detectors, often require cryogenic or thermoelectric cooling to maintain laser performance and minimise thermal noise at the detector side. Furthermore, the high cost of MIR sources and cryogenically cooled detectors restricts accessibility and limits widespread adoption of mid-IR spectroscopy to well-funded research labs [6]. Collectively, these challenges highlight the need for alternative MIR spectroscopic sensing and imaging approaches that offer enhanced sensitivity, accuracy, and cost-effectiveness while maintaining compact and robust device designs, which are important for biological and biomedical applications.

Quantum sensing methods are gaining popularity as they offer an alternative that combines competitive sensitivity with reduced reliance on specialised MIR sources and detectors. Quantum technologies based on entangled photon pairs [7-16] provide new opportunities for quantum MIR sensing and imaging. These are typically based on the enabling process of spontaneous parametric down-conversion (SPDC) in which a single high energy pump photon interacts with a second order ($\chi^2$) nonlinear crystal, spontaneously producing an entangled photon pair of lower energy photons, known as signal and idler [8, 9]. Due to the energy and momentum conservation laws between pump, signal and idler, the entangled pair is correlated in both frequency and time. A variety of quantum spectroscopy with undetected photon schemes has been demonstrated for gases, solid state and liquid samples in the mid-infrared fingerprint region [6, 9-13, 17]. However, this approach has not yet been extended to liquid-phase biomolecular mixtures such as protein solutions, where strong water absorption and complex vibrational structure present additional challenges.

In this work, we evaluate the feasibility of applying Quantum Spectroscopy with Undetected Photons (QSUP) to the detection of proteins in aqueous solution. Using FTIR measurements of well-characterised proteins as input, we model QSUP visibility in the amide I–II range and map the technique's parameter space (including crystal length, sample thickness, and interferometer geometry) required for biomolecular sensing. The benefits of the proposed QSUP are twofold: first, it eliminates the need for MIR lasers and detectors by using widely available and cost-effective equipment operating in the visible range; second, it identifies parameter regimes where QSUP can, in principle, achieve sensitivity for protein detection comparable with conventional FTIR. In this work, we optimise the QSUP pair-generation method to achieve high absorption sensitivity, while providing clear guidance for data interpretation. By connecting the spectral transmissivity of a biomolecular sample with the entangled photon pairs interference pattern, we show that information conventionally obtained by MIR instrumentation can be accessed using visible/NIR lasers and detectors, enabling a path toward compact, low-noise, and cryogen-free spectroscopic systems.

## 2. Principles of Quantum Spectroscopy with Undetected Photons

Quantum spectroscopy with undetected photons (QSUP) originates from early advances in nonlinear optics, including the first demonstrations of SPDC in the late 1960s and early 1970s, which established the generation of correlated photon pairs for quantum measurements [18]. The core principle enabling QSUP, induced coherence without induced emission, was introduced in the 1990s. This first demonstration showed that interference can be observed even when one of the photons is never detected [19]. This concept was experimentally transformed into a practical imaging technique in 2014, when quantum imaging with undetected photons demonstrated that spatial information carried by IR photons can be retrieved using only visible range detection, without coincidence counting [20]. The technique was later extended to spectroscopy, confirming that both amplitude and phase information of a sample interrogated by undetected photons can be encoded into the interference of the detected partner photon, enabling spectroscopy and refractive index imaging without direct MIR detection [11, 21].

QSUP has since evolved into a powerful spectroscopy and microscopy method, particularly for spectral regions where detectors are inefficient, expensive, or require cryogenic cooling. By mapping the spectral signatures of undetected MIR photons onto visible or NIR photons, the technique enables chemically specific measurements using low-noise silicon detectors [22]. Recent demonstrations have shown that QSUP and its imaging counterpart can probe a wide range of sample types, including gaseous molecular absorption (e.g., $CO_2$ and nitrous oxide) [11, 12], liquids [22], solid-state and polymer samples [13, 23], biological tissues [24] and varies phase-contrast objects [25]. Recently, QSUP has been extended into full quantum Fourier-transform infrared (QFTIR) schemes, demonstrating broadband spectroscopy of gases and solids in the fingerprint region with competitive signal-to-noise performance relative to classical FTIR for a given MIR probe intensity [10, 16, 26]. Complementary work on quantum IR attenuated total reflection (QIR-ATR) has shown that undetected-photon architectures can also address strongly absorbing media such as aqueous samples by exploiting evanescent-field interactions at an interface [27]. Despite these advances, QFTIR and QIR-ATR face practical limitations. QFTIR implementations typically rely on Michelson-type interferometers and dual-crystal SPDC sources, which increase instrument complexity, acquisition time, while setting higher demand on instrument's stability [10, 16]. In contrast, QIR-ATR relies on evanescent-field sensing at the crystal–sample interface, leading to a significantly reduced interaction volume and consequently lower sensitivity for dilute liquid-phase samples compared to bulk transmission measurements [27].

At the same time, access to protein secondary structure in native, liquid environments are central to understanding folding, conformational transitions, and biomolecular function. In practice, protein biomolecules are often available only at low concentrations and must be measured in aqueous buffers, where the strong and broadband mid-infrared absorption of water severely restricts measurements at low protein concentrations [4]. These challenges motivate the development of alternative spectroscopic strategies that can operate under realistic bioanalytical conditions while retaining sensitivity to subtle structural changes. To date, QSUP has not been applied to liquid-phase proteins in aqueous buffer, nor has it been linked to quantitative secondary-structure analysis. In the following, we address this challenge using a double-pass QSUP configuration specifically tailored to the amide I–II bands and demonstrate through numerical modelling how optimisation of the interferometer parameters enables required sensitivity to detect protein secondary-structure signatures in aqueous solution.

The central idea of QSUP is illustrated in Fig. 2 (a). A continuous wave (CW) pump laser passes through a nonlinear crystal, which generates entangled pairs at signal ($\lambda_s$) and idler ($\lambda_i$) wavelengths via SPDC process. The MIR idler passes through the sample to be measured. Both the pump and the generated SPDC photons are reflected by a flat mirror located at the focal plane of a 90° off-axis parabolic mirror (OAPM). Upon reflection all photons travel back through the crystal for a second time, where the pump generates additional photon pairs. The newly generated signal photons interfere with the signal photons from the first pass, and the interference pattern is measured by a visible/NIR detector. The OAPM also performs an inverse Fourier transform on the returning photons, restoring their original transverse profile. While the flat mirror remains fixed, the nonlinear crystal can be translated by a small distance relative to its initial position, enabling the adjustment of the visibility of the resulting interference fringes.

For non-collinear SPDC process, the signal intensity, $I_s$, on the first pass through the nonlinear crystal depends on a phase-matching condition [6, 9], which in turn depends on the wavelength ($\lambda_s$) and angle ($\theta_s$) of the signal beam:

$$I_s(\lambda_s, \theta_s) \propto C_0 [\text{Sinc}(\frac{\delta}{2})]^2. \qquad (1)$$

We chose x-axis along the pump propagation and z-axis is transverse (vertical) direction to x. The longitudinal phase mismatch accumulated over the nonlinear crystal length is $\delta(\lambda_s, \theta_s) = [k_p - k_s \cos \theta_s - k_i \cos \theta_i] L$ with $k_p, k_s$ and $k_i$ as the pump, signal, and idler wave vectors and L is the crystal length. The transverse momentum component determines the idler angle as $\theta_i(\lambda_s, \theta_s) = \arcsin[(k_s/k_i) \sin \theta_s]$. These relations reflect the underlying momentum conservation governed by SPDC process (see Fig. 2 (b)) with the efficiency $C_0$ expressed as [9]

$$C_0 = \frac{16}{3}\left(\frac{c\pi^3 \hbar L^2 P_p d_{eff}^2}{\varepsilon_0 n_p n_s n_i \lambda_s^3 \lambda_i S_{eff}}\right), \quad (2)$$

where c is the speed of light, $\hbar$ is the reduced Planck constant, $\varepsilon_0$ is the vacuum permittivity, and $n_p$, $n_s$ and $n_i$ are the refractive indices for the pump, signal, and idler respectively. $S_{eff}$ represents the effective cross-sectional area of the interacting beams and $d_{eff}$ is the nonlinear coefficient of the crystal. Most of the crystal parameters, such as refractive indices, are fixed, but the length of the nonlinear crystal L can be chosen to optimise phase matching $\delta(\lambda_s, \theta_s)$ and improve SPDC efficiency $C_0$. In addition, the pump, signal, and idler photon wavelengths are related through energy conservation, such that the pump photon energy is equal to the sum of the signal and idler photon energies (see Fig. 2 (C)).

When a medium absorbing at the idler wavelength is introduced into the double-pass setup, the interference pattern of the signal photons is modified, and is given by [12]

$$I_s(\lambda_s, \theta_s) \propto C_0 [\text{sinc}(\tfrac{\delta}{2})]^2 [1 + |\tau|\cos(\delta + \delta_s)] \quad (3)$$

The additional term, $C_0|\tau|\cos(\delta + \delta_s)[\text{sinc}(\tfrac{\delta}{2})]^2$, accounts for the modulation caused by signal interference. $\delta_s$ is the additional phase mismatch introduced by the sample, the biocell, and any air gaps present in the optical path. In the double-pass configuration, this modulation depends on the phases acquired by the pump, signal, and idler photons as they propagate through the nonlinear crystal twice and through the medium in the idler path. If the medium absorbs idler light, the visibility V (i.e. contrast between maxima and minima of interference fringes) of the signal interference pattern is reduced. This can be directly linked to the sample's transmissivity ($\tau$) in the MIR wavelength range [12]:

$$V = \frac{I_s^{max} - I_s^{min}}{I_s^{max} + I_s^{min}} = \tau \quad (4)$$

In addition to crystal length L, other key parameters such as sample thickness $L_m$, the distance between the nonlinear crystal and the OAPM $L_a$, and the biocell thickness $L_b$ significantly influence the interference contrast.

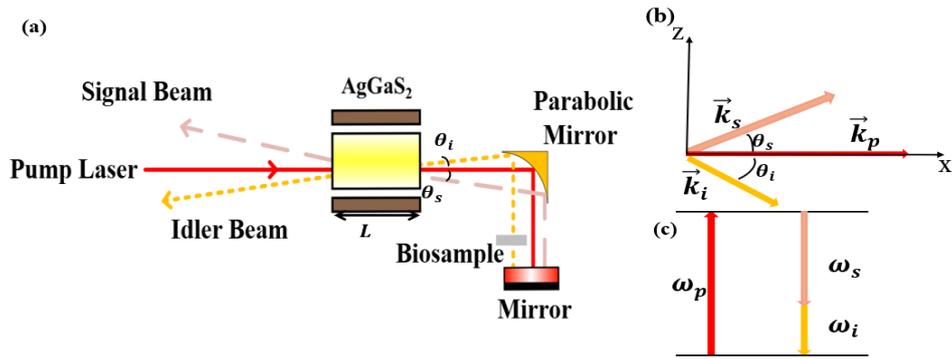

**Figure 2.** SPDC interferometer configuration and phase-matching conditions. **(a)** Double-pass SPDC configuration consisting of a nonlinear crystal, an off-axis parabolic mirror (OAPM), and a flat mirror, **(b)** Illustration of transverse and longitudinal momentum conservation in the non-collinear SPDC process, **(c)** Energy conservation diagram showing the relation between pump, signal, and idler photon frequencies.

## 3. Sample preparation and FTIR measurement

Lyophilized bovine serum albumin was purchased from Merck (Bayswater, Victoria) and Recombinant Human NT-proBNP was purchased from Millennium science (Mulgrave, Victoria), both proteins were used as supplied. Phosphate buffer solution (PBS) was purchased from ThermoFisher Scientific (Scoresby, Victoria). Protein solutions were prepared by dissolving individual proteins in PBS at 3 mg/ml which is the limit of detection of the FTIR instrument.

FTIR spectra of the protein solutions were collected using a Bruker BioATR spectrometer. For each measurement, 15 µL of sample was loaded into the BioATR reservoir. Spectra were recorded at a resolution of 4 cm$^{-1}$, with 64 scans averaged per spectrum. Measurements also were recorded at four temperatures, 24 °C, 36 °C, 55 °C and 68 °C to investigate temperature-dependent relative changes in protein secondary structure, such as α-helices and β-sheets. Prior to acquisition, each sample was equilibrated for 5 min at the target temperature.

FTIR results for the BSA and NT-proBNP samples are shown in Fig. 3 (a), highlighting differences in molecular vibrations and characteristic absorption bands. Temperature-dependent spectral changes in BSA are presented at 24 °C and 68 °C in Fig. 3 (b), indicating structural alterations induced by heating. Vertical dotted lines mark the wavelength positions of the Amide I band maxima for each case. The horizontal dashed line at the bottom of the plot emphasizes the relative spectral shift. Although the absolute spectral shift with temperature is small, it corresponds to significant changes in the secondary structure of the Amide I band. Small shifts in this region are commonly associated with temperature-induced modifications of hydrogen bonding and relative contributions of the secondary structure elements such as α-helices, β-sheets, and disordered structures [5]. A more comprehensive discussion is provided in the Supplementary Information.

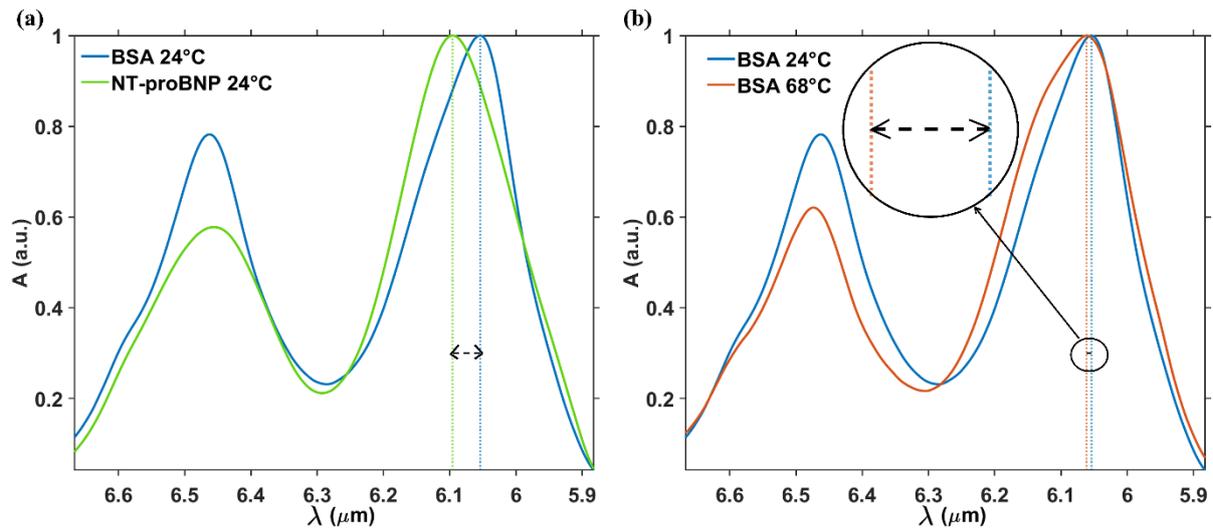

**Figure 3.** FTIR results **(a)** Comparison of the FTIR spectra of BSA and NT-proBNP samples, highlighting differences in their molecular vibrations and characteristic bands. **(b)** Temperature-dependent changes in BSA spectra observed at 24 °C and 68 °C, showing structural modifications with heating.

The changes in the secondary structure of proteins were analysed after subtraction of atmospheric water vapor and carbon dioxide contributions from the IR spectra, followed by baseline correction. The positions of individual component peaks were determined from the second derivative spectra (Savitzky-Golay, 9 smoothing points) and the baseline corrected Amide I band. Peak assignments to specific secondary structures were made based on the previous report [4]. Quantification was performed by fitting Gaussian line shape to each peak using OPUS 8.7.31 (Bruker, OPUS software) and integrating the peak area. The latter are expressed as a percentage of the total Amide I band area, representing the relative proportion of each secondary structure component. For NT-proBNP at 24 °C, the amide I region was divided into three distinct components (Table 1). The most prominent band at 1638 cm$^{-1}$ was assigned to α-helix structures, which accounts for ~82% of the total secondary structure. A minor contribution from β-turn structures was observed at 1662 cm$^{-1}$ (~1%), while β-sheets at 1686 cm$^{-1}$ represented ~17% of the structural content. Compared with BSA, NT-proBNP exhibited higher α-helix and β-sheet content at the same 24 °C temperature while having a reduced β-turn contribution.

Table 1. Amide I band frequencies, corresponding secondary structure assignments and their relative percentage occurrence for BSA (3mg/mL in PBS) at 24 °C and 68 °C, to assess temperature-induced structural changes, and NT-proBNP.

|  | Position (cm$^{-1}$) | Secondary Structure | % |
|---|---|---|---|
| **BSA 24 °C** | 1627±4 | β- sheet | 8.6 |
|  | 1655±4 | α- helix | 78.7 |
|  | 1683±4 | β- turn | 12.7 |
| **BSA 68 °C** | 1626±4 | β- sheet | 26.4 |
|  | 1655±4 | α- helix | 64.5 |
|  | 1680±4 | β- turn | 9.1 |
| **NT-proBNP 24 °C** | 1638±4 | α- helix | 81.9 |
|  | 1662±4 | β- turn | 0.6 |
|  | 1687±4 | β- sheet | 17.5 |

## 4. QSUP simulations

We now consider the use of QSUP for label-free biomolecular detection and dynamic protein structure studies in the MIR. The intensity distribution of the signal beams interference in the presence of a biological sample is given by equation (3). Here we use the ATR-FTIR measurement data for the sample transmissivity τ to test feasibility of the QSUP approach. An important advantage of QSUP compared to ATR-FTIR measurement is that it uses a transmission approach where the MIR idler photons probe the sample across its entire length. In contrast, ATR-FTIR measurements rely on evanescent-field sensing and are inherently surface sensitive. We consider a CW 660 nm laser with an output power of 100 mW as the source of entangled photon generation. The idler wavelengths lie in the MIR range of 5.88–6.66 μm corresponding to the Amide I and II vibrational bands of biomolecules. Based on energy conservation in the SPDC process (see Fig. 2 (c)), this corresponds to signal photons in the visible/NIR range of 732–743 nm. The phase-matching condition, in turn, sets a requirement for the choice of a nonlinear crystal, which needs to be transparent at the pump, signal, and idler wavelengths simultaneously. We therefore selected AgGaS$_2$ (or AGS) due to its broad transparency window (0.53–12 μm) [12]. When the crystal is cut at 74°, type-I phase matching is achieved for all selected wavelength bands with $d_{eff}$ = 15.5 pm/V. Phase matching was first verified in the absence of a sample, using the selected pump and idler wavelengths, confirming the feasibility of the AgGaS$_2$ crystal to produce entangled pairs at the appropriate wavelengths. Next, the FTIR data for 3 mg/mL BSA at 24 °C was incorporated into the simulations to model the signal angular–wavelength interference pattern. The result is shown in Fig. 4 (a). Signal photons are emitted primarily within ±1° cone centred at the direction of the pump beam. Integrating $I_s(\lambda_s, \theta_s)$ over the emission angle $\theta_s$ collapses the 2D interference map into a 1D spectrum $I_s(\lambda_s)$, representing the total emitted power at each wavelength as a spectrometer would measure (see Fig. 4 (b)). Variations in the maxima and minima of the interference pattern $I_s(\lambda_s)$ represents its visibility as described by equation (4) and it related to the sample absorption (see Fig. 4 (c)). Visibility values are extracted from adjacent peak–dip pairs of $I_s(\lambda_s)$ and assigned to the midpoint wavelength between each peak and dip. Horizontal error bars denote the spectral separation of the corresponding peak–dip pairs. We also can define a weighted visibility parameter $\beta = V_A(V_A - V_B)$, which will be used in the next section. In general, we aim to maximise $\beta$ as this corresponds to high overall visibility of the interference pattern ($V_A \to 1$) with clear differentiation between the regions of interest, with and without absorption ($V_A - V_B$). This process is especially important if the instrumentation used, e.g. a camera, does not have the highest performance (average or low quantum efficiency, high dark current noise etc.) Therefore, the parameter $\beta$ is introduced to describe a condition that is optimal in two respects, a high detected photon count rate and a large change in visibility simultaneously.

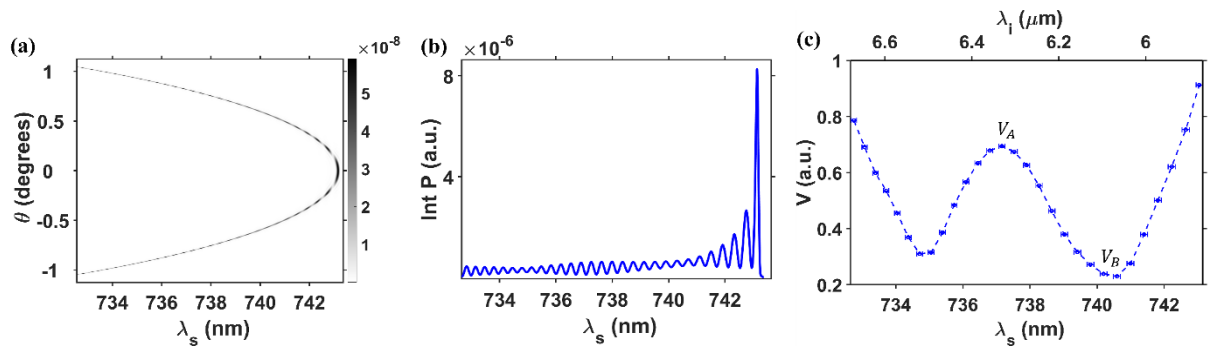

**Figure 4.** Workflow for QSUP method: **(a)** Simulation of the wavelength–angular interference pattern of signal photons in the presence of the 3mg/ml BSA at 24 ̊C. **(b)** 1D signal interference across the phase-matched 733-743 nm wavelengths region, obtained by angle integration of **(a)**. **(c)** The visibility map of the interference fringes reflects the sample transmissivity in the MIR range.

## 4.1. Optimization of parameters for QSUP

Several experimental parameters play a central role in determining the weighted visibility β and spectral response of QSUP technique. In particular, the nonlinear crystal length (L), the sample length ($L_m$), and the separation between the nonlinear crystal and the parabolic mirror ($L_a$) are all important to the overall outcome of QSUP detection. The numerical model we developed is well-suited to identify parameter regimes that maximise visibility.

First, we investigate the influence of the nonlinear crystal length and show β as a function of L (Fig. 5 (a)). A longer crystal length L leads to higher number of entangled photon pairs being generated in the crystal, improving the signal-to-noise level of the technique. On the other hand, a longer L leads to a narrower phase-matching bandwidth due to accumulated walk-off between the pump, signal and idler photons, which can make the fringe detection more difficult and sets a requirement for a high-resolution spectrograph. Thus, there is a trade-off between SPDC efficiency and spectral resolution, expressed by rise and fall of β(L) curve as shown in Fig. 5 (a). It can be seen that as L increases, β initially does so as well, reflecting enhanced SPDC efficiency and improved fringe contrast due to the higher photon-pair flux. However, for the crystal lengths greater than 6 mm, β starts to reduce reflecting a decline in fringe detectability. Overall, there is minimal difference between L = 5 and 6 mm and both can be identified as optimal in this study. Since longer crystals involve greater fabrication complexity and cost, a length of L = 5 mm is chosen as the optimal for the proposed QSUP configuration.

Next, we examine the influence of the sample length $L_m$ as illustrated in Fig. 5 (b). Recall that $L_m$ determines the interaction strength between the MIR idler photons and the biomolecular sample. As expected, increasing $L_m$ leads to stronger absorption and a reduction in visibility. Therefore, initially increasing $L_m$ leads to a rise of β due to the increase in the term ($V_A - V_B$), reflecting improved interaction between the MIR idler and the sample. Beyond an optimal sample length of $L_m = 6$ μm, however, β starts to roll off as the peak visibility $V_A$ drops down as well. Thus, a sample length of 6 μm is selected as the optimal value in this study.

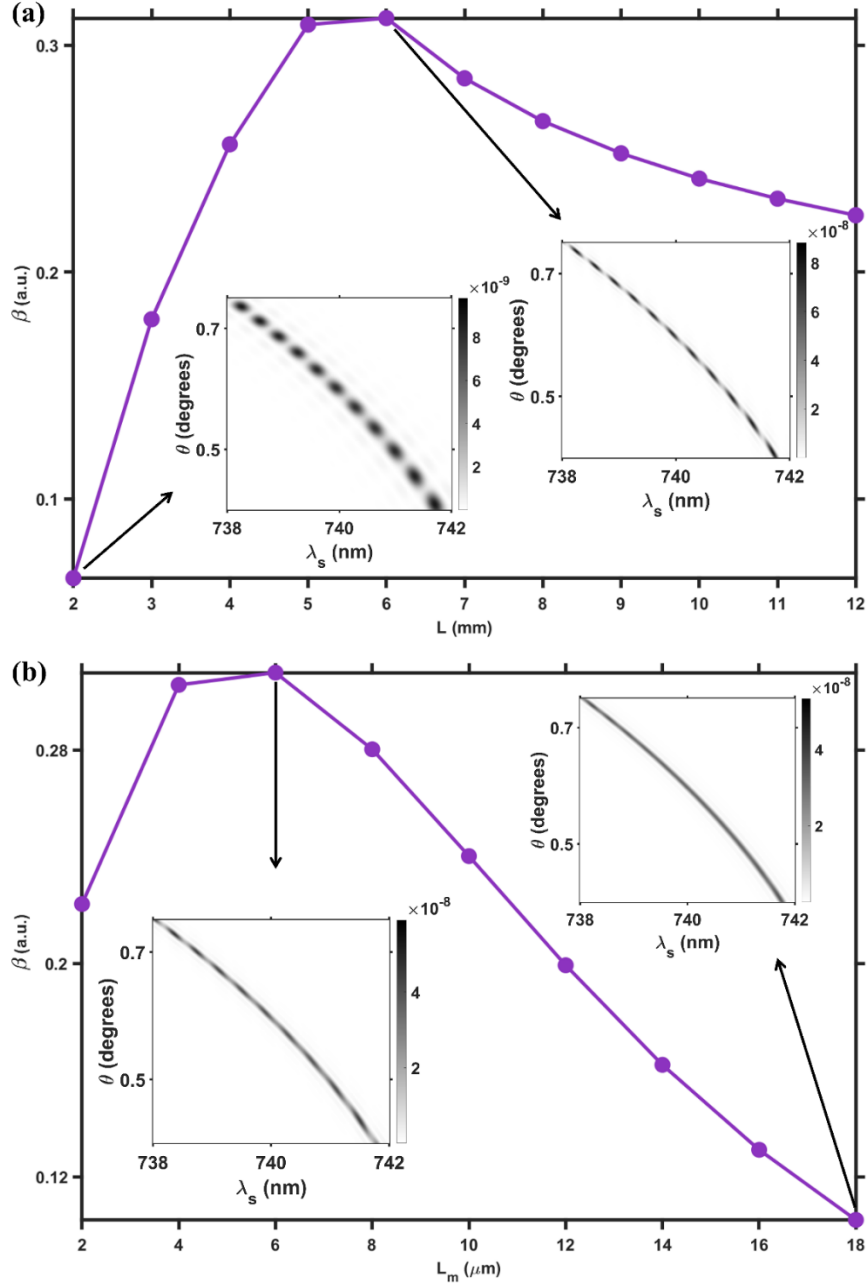

**Figure 5.** Optimisation of experimental parameters using weighted visibility β: **(a)** Weighted visibility β as a function of the nonlinear crystal length L, showing an optimal range of 5–6 mm. **(b)** The weighted visibility β for a varying optical path length of the sample $L_m$ identifies the optimal sample length $L_m = 6$ μm. For each case, the corresponding interference fringe patterns at the maximum and minimum values of β are also shown.

We finally examine the role of the crystal–mirror separation $L_a$, as shown in Fig. 6. Varying $L_a$ introduces wavelength-dependent phase shifts between the two SPDC passes, modifying the conditions for constructive and destructive interference. As a result, changes in $L_a$ lead to the interference fringe tilting and changes in the number of fringes over the wavelength range of interest, with consequences to fringe visibility estimation. By appropriately tuning this distance, the interference fringes can be aligned to maximise contrast, enabling fine control over the spectral response without altering the overall optical power. The crystal-mirror separation of $L_a$ = -8.75 mm yields the optimal case for which the fringes are well separated, and their contrast is the highest, thus resulting in the maximised β. However, we observe an anomaly in the β($L_a$) curve around $L_a \approx -7$ mm, where the visibility approaches zero. In this region, the linear interference term $\cos(\delta + \delta_s)$ in the equation (3) is constant over the signal wavelength of interest and the fringes vanish, meaning that the visibility of the fringes can no longer be calculated. It can be shown (see Supplementary information) that for a near-colinear SPDC process the phase accumulation along the air gap exactly cancels the phase from the air gap. The length of the air gap where this happens is

$$L_a^{min} = -\frac{1}{n_b} L_b \quad (5)$$

where $n_b$ is the refractive index and $L_b$ is the total path length through the biocell. For the situation in Fig. 6, for the biocell index of $n_b = 1.4324$ and path length $L_b = 10$ mm we obtain $L_a^{min} = -6.98$ mm, which coincides with the observed minimum. Because the air gap can be modified during the experiment, this gives an indication of which regime to avoid when performing the QSUP experiments.

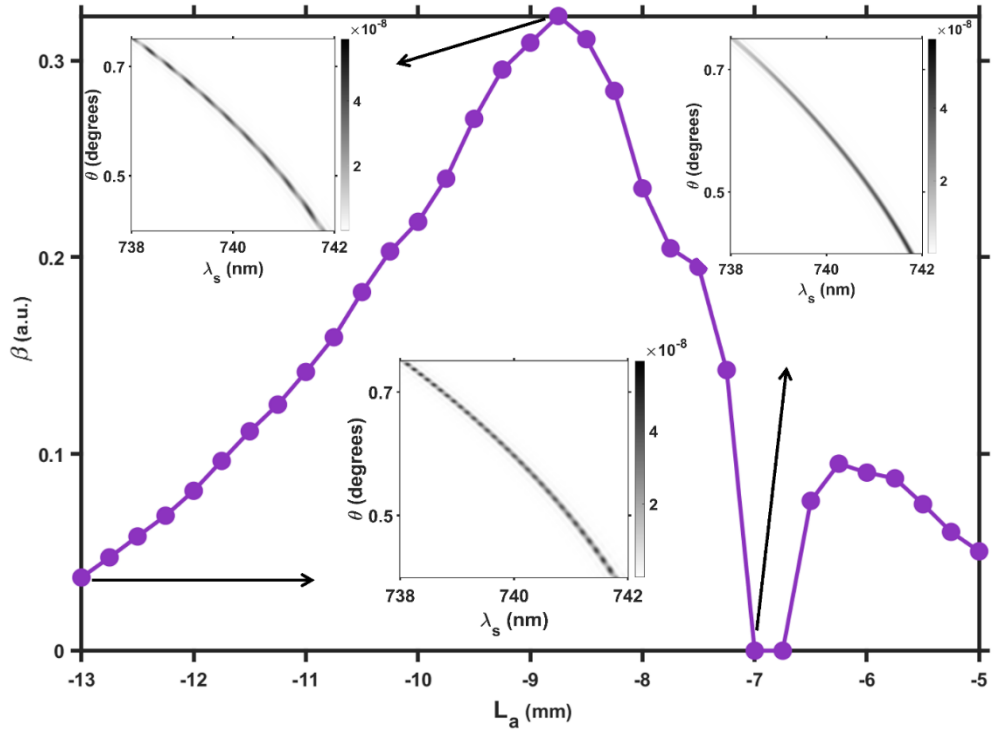

**Figure 6.** The weighted visibility β for a varying crystal–mirror separation $L_a$ indicates that the optimal separation is $L_a = -8.75$ mm.

## 4.2. Application of QSUP to biomolecular sensing of proteins

Next, we tested the QSUP numerical model on the two protein samples, the BSA and NT-proBNP. To evaluate the feasibility of this technique we used experimentally measured MIR absorption spectra for each of the samples, obtained using a standard FTIR technique (see Fig. 7). Results demonstrate a good correlation between the visibility retrieved from the quantum spectroscopy setup and the amplitude transmissivity derived from classical FTIR measurement. Higher visibility indicates greater transmission (lower absorption) within the region of interest. In Fig. 7 (a), both protein samples exhibit distinct visibility minima corresponding to their molecular absorption bands, Amide I and II peaks. Figure 7 (b) compares BSA spectra at 24 °C and 68 °C, showing a shift in the curve minima corresponding to Amide I peak with increasing temperature and an increase in visibility in the Amide II region. These spectral changes indicate structural alterations associated with protein denaturation, where heating disrupts the secondary structure and modifies the vibrational response of the amide bands.

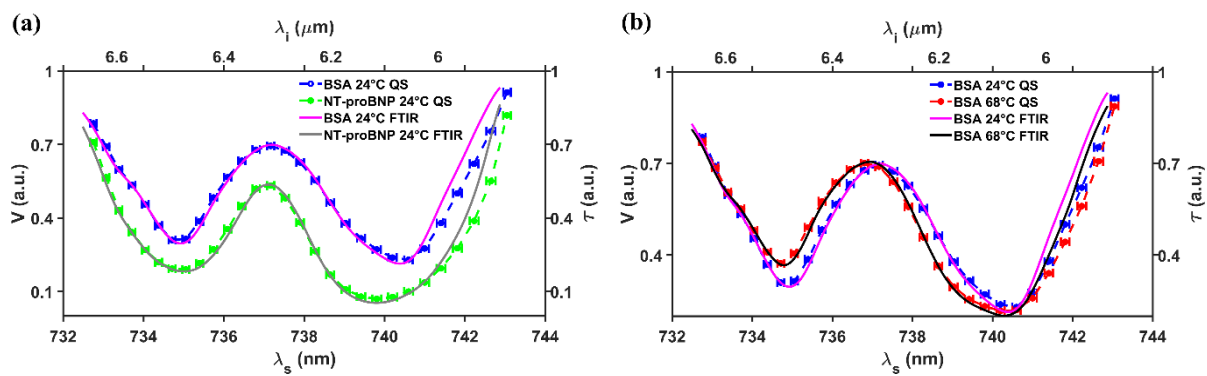

**Figure 7.** Comparison between the visibility retrieved from the quantum spectroscopy simulation with 6 µm-long sample and the FTIR transmission data. **(a)** Comparison between BSA and NT-proBNP protein samples. **(b)** Temperature-dependent changes in BSA spectra at 24 °C and 68 °C.

## 5. Discussion and Conclusion

Our results demonstrate a close agreement between QSUP and FTIR results for the two standard proteins and can, in principle, be extended to sensing of other biomolecules in aqueous solutions. By embedding experimentally measured FTIR spectra of BSA and NT-proBNP into a numerical model of a double-pass interferometer, we numerically obtained visibility spectra that closely follow the classical transmissivity in the amide I–II region and resolve temperature-induced changes in protein secondary structure. This establishes that undetected-photon spectroscopy can access biomolecular structural information under sample conditions compatible with standard bioanalytical workflows, while relying exclusively on visible-range sources and detectors [11-13, 20-22]. A central practical advantage of the proposed approach is the use of a continuous-wave 660 nm pump laser and VIs/NIR single-photon detectors. Lasers at this wavelength are widely available, cost-effective, and highly stable, while silicon-based detectors offer high quantum efficiency, low dark noise, and room-temperature operation. This removes the need for specialised mid-infrared sources and cryogenically cooled detectors that are typically required in conventional MIR spectroscopy [6], providing a simplified and potentially lower-cost experimental platform [12].

In contrast to earlier demonstrations of infrared spectroscopy with visible light and broadband QFTIR, largely focused on gases, solids, and simple liquids [10-13, 16, 26] the present work addresses proteins in liquid solutions and directly links quantum observables to established secondary-structure analysis. The protein concentration used here (3 mg/mL) is realistic for ATR-FTIR and near practical detection limits in aqueous environments [4]. Importantly, the transmission-based QSUP design requires smaller sample volumes than conventional ATR-FTIR: while ATR typically uses ≳15 µL, QSUP achieves reasonable sensitivity with only ~10 µL in a 6 µm long biocell chamber. In addition, QSUP probes the sample in transmission, accessing bulk absorption over a defined path length, unlike the fast-decaying evanescent field of ATR-FTIR that has a penetration depth of typically only a few hundreds of nanometres. The FTIR-informed modelling further allows us to identify operating regimes that optimise visibility contrast in the amide I–II bands as a function of crystal length, sample thickness, and interferometer geometry, thereby providing quantitative design rules for future implementations and possible improvements to the detection sensitivity and specificity.

This work focuses on the numerical modelling and theoretical framework of QSUP, rather than a full experimental demonstration. Future efforts will target real-time measurements, noise characterisation, and determination of detection limits. In SPDC-based setups, pump stability, interferometer drift, and photon-pair brightness directly affect fringe visibility and achievable SNR, particularly for weakly absorbing samples [10]. Realising the full potential of QSUP for protein sensing will therefore require experimental implementations of the proposed geometry, systematic benchmarking against state-of-the-art FTIR under identical conditions, and extensions to lower concentrations and flow-cell or microfluidic sample delivery.

The methodology developed here is readily generalised. While the present study considers AgGaS$_2$ pumped at 660 nm as one practical implementation, alternative nonlinear materials and pump wavelengths can be employed to extend spectral coverage across broader regions of the MIR and into the far-infrared and terahertz domains, where low-energy collective and intermolecular modes of biomolecules reside [7, 17, 28]. Beyond absorption spectroscopy, the intrinsic phase sensitivity of induced-coherence interferometry also enables refractive-index retrieval and combined absorption–dispersion measurements [11, 12], opening pathways to quantum-enabled studies of protein conformational changes and interfacial phenomena.

Finally, integrating the source and interferometer into compact, fibre-coupled or chip-scale platforms [29], and combining them with established liquid-handling and temperature-control hardware, could enable robust, room-temperature instruments for MIR biomolecules detection. We note that significant challenges in implementation of this method for field work, especially for applications requiring broadband operation; this is primarily due to the stringent requirements for optical alignment of the quantum interferometer within the short coherence length.

In conclusion, we have presented a numerical framework demonstrating that QSUP architecture can reproduce key MIR absorption features of proteins in aqueous environments using visible/near-infrared sources and detectors. Through systematic optimisation of parameters such as the nonlinear crystal length, sample thickness, and interferometer geometry, we identified the optimal regime for photon-pair generation efficiency and maximised interferometric visibility. While the present results are based on numerical modelling, they indicate that QSUP can provide access to MIR spectroscopic information under conditions compatible with standard bioanalytical environments. Importantly, this capability is highly relevant for medical and biomolecular sensing, where both the detection of protein biomarkers and the assessment of protein structural signatures in the amide I-II region can provide diagnostically meaningful information. Future work will focus on experimental implementation, noise characterisation, and benchmarking against established FTIR techniques to assess further QSUP sensitivity limits and its practical performance.

## 6. Data availability statement

All the data supporting this study are available upon request to the corresponding authors.

## 8. Funding Declaration

The authors acknowledge the financial support from Australian Research Council Centre of Excellence in Quantum Biotechnology (CE230100021) and IRTP from UTS.

## 9. Author contribution statement

## 10. Additional Information

## 11. Figure legends

**Figure 1.** Conceptual scheme for retrieving mid-infrared (MIR) spectroscopic information of proteins using visible (VIS) / near-infrared (NIR) photons. A liquid biopsy sample containing proteins exhibits characteristic vibrational absorption features in the MIR spectral region, such as the Amide I and Amide II bands. In the proposed approach, correlated photon pairs are generated through spontaneous parametric down-conversion (SPDC), producing an idler photon in the MIR range and a signal photon in the VIS/NIR range. The MIR idler photon interacts with the protein sample and acquires information about its absorption. Through quantum correlations between the photon pairs, this interaction modifies the interference properties of the detected signal photons. As a result, the MIR spectroscopic response of the sample can be retrieved by measuring the VIS/NIR photons with conventional detectors, enabling indirect access to molecular vibrational information without requiring direct MIR detection.

**Figure 2.** SPDC interferometer configuration and phase-matching conditions. **(a)** Double-pass SPDC configuration consisting of a nonlinear crystal, an off-axis parabolic mirror (OAPM),

and a flat mirror, **(b)** Illustration of transverse and longitudinal momentum conservation in the non-collinear SPDC process, **(c)** Energy conservation diagram showing the relation between pump, signal, and idler photon frequencies.

**Figure 3.** FTIR results **(a)** Comparison of the FTIR spectra of BSA and NT-proBNP samples, highlighting differences in their molecular vibrations and characteristic bands. **(b)** Temperature-dependent changes in BSA spectra observed at 24 °C and 68 °C, showing structural modifications with heating.

**Figure 4.** Workflow for QSUP method: **(a)** Simulation of the wavelength–angular interference pattern of signal photons in the presence of the 3mg/ml BSA at 24 ˚C. **(b)** 1D signal interference across the phase-matched 733-743 nm wavelengths region, obtained by angle integration of **(a)**. **(c)** The visibility map of the interference fringes reflects the sample transmissivity in the MIR range.

**Figure 5.** Optimisation of experimental parameters using weighted visibility β: **(a)** Weighted visibility β as a function of the nonlinear crystal length L, showing an optimal range of 5–6 mm. **(b)** The weighted visibility β for a varying optical path length of the sample $L_m$ identifies the optimal sample length $L_m = 6$ μm. For each case, the corresponding interference fringe patterns at the maximum and minimum values of β are also shown.

**Figure 6.** The weighted visibility β for a varying crystal–mirror separation $L_a$ indicates that the optimal separation is $L_a = -8.75$ mm.

**Figure 7.** Comparison between the visibility retrieved from the quantum spectroscopy simulation with 6 μm-long sample and the FTIR transmission data. **(a)** Comparison between BSA and NT-proBNP protein samples. **(b)** Temperature-dependent changes in BSA spectra at 24 °C and 68 °C.

**Table 1.** Amide I band frequencies, corresponding secondary structure assignments and their relative percentage occurrence for BSA (3mg/mL in PBS) at 24 °C and 68 °C, to assess temperature-induced structural changes, and NT-proBNP.

# Quantum Spectroscopy with Undetected Photons for Biomolecular Sensing in the Mid-Infrared: Supplementary Information

Mahya Mohammadi[1,2*], Meryem-Nur Duman[1,2], Isa Ahmadalidokht[1,2], Mohammad Sadraeian[2,3], Christopher G. Poulton[1], Alexander S. Solntsev[1,2], Irina V. Kabakova[1,2*]

[1.] School of Mathematical and Physical Sciences (MaPS), Faculty of Science, University of Technology Sydney, Sydney, NSW 2007, Australia

[2.] ARC Centre of Excellence in Quantum Biotechnology (QUBIC), University of Technology Sydney, Sydney, NSW 2007, Australia

[3] Institute for Biomedical Materials and Devices (IBMD), Faculty of Science, University of Technology Sydney, Sydney, NSW 2007, Australia

*Correspondence to: Mahya.Mohammadi@student.uts.edu.au, Irina.Kabakova@uts.edu.au

## SUPPLEMENTARY NOTE 1: FTIR Measurements

The FTIR spectra used in this work were acquired using a BioATR FTIR instrument, while the QSUP model assumes light transmission through a sample. To relate these two descriptions, the ATR absorbance spectrum was used to estimate the effective absorption coefficient of the sample. In ATR spectroscopy the infrared field interacts with the sample through an evanescent wave with a penetration depth depending on the light wavelength and the refractive indices of the ATR crystal and sample. Using the known refractive index of the ZnSe ATR crystal and the measurement geometry, the average penetration depth in MIR range was estimated to be 100 nm and used to convert the measured ATR absorbance into an absorption coefficient. This coefficient was then used to construct an effective transmissivity amplitude $\tau$ for the QSUP model via the Beer–Lambert relation [1]. This approach allows us to compare directly ATR measurements with QSUP numerical simulations.

### A. Secondary structure calculations

The negative normalized second derivative of the BSA absorbance at 24 °C, fitted with Gaussian line shape to resolve secondary structure features, is shown in Supplementary Fig. S1 (a). The absorbance is also converted into transmissivity amplitude, which serves as the primary parameter for subsequent analysis in this work. Supplementary Fig. S1 (b) presents the absorbance spectrum of BSA at 24 °C alongside its corresponding transmittance.

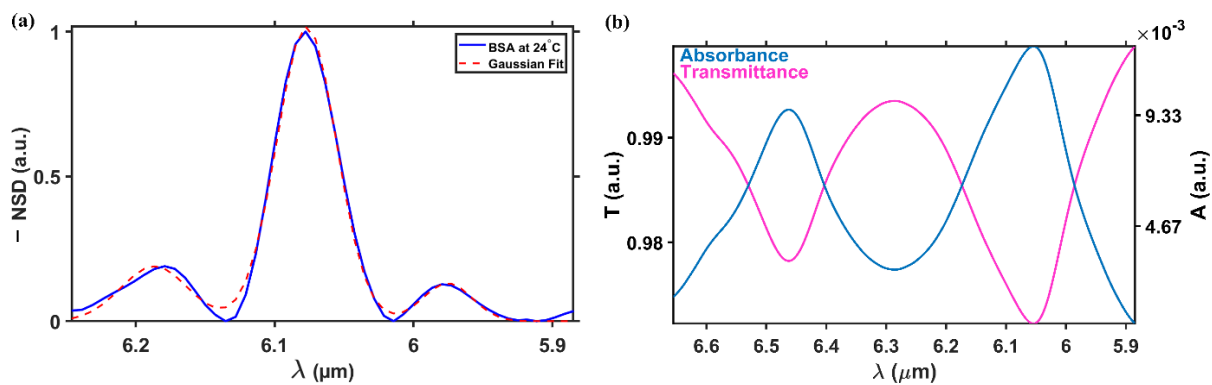

**Supplementary Figure S1**. **(a)** Analysis of the minus normalized second derivative of BSA absorbance at 24 °C, with Gaussian curve fitting applied to identify secondary structure components. **(b)** Comparison between the absorbance and transmissivity amplitude of BSA at 24 °C

## B. Temperature dependent structural changes of BSA

Elevated temperatures are known to induce protein denaturation and alter secondary structure. The amide I band, which contains information on α-helix, β-sheet, β-turn, antiparallel intermolecular β-sheet and random coil [2] is commonly used to analyse and quantify protein secondary structure. In this study, FTIR spectroscopy was used to investigate temperature dependent structural changes in BSA and associated protein dynamics [3].

Our results confirm that the secondary structure of BSA is sensitive to temperature. Figure 3 **(b)** shows the FTIR spectra of BSA at 24 °C and 68 °C, with a clear change at the elevated temperature indicated by the appearance of a shoulder within the amide I region. This spectral feature suggests a redistribution of secondary structure components. Following second derivative analysis and Gaussian curve fitting (see Supplementary Fig. **S**1 (a)), individual bands within the amide I region were assigned (**Table 1**) based on previous studies [4-7]. At 24 °C, the band at 1655 cm$^{-1}$ was attributed to α-helix (~79 %) [4], while the bands at 1627 and 1683 cm$^{-1}$ corresponded to β-sheet (~9%) and β-turn (~13%) [4, 8]. At 68 °C, the α-helix and β-turn content decreased to ~65 % and ~9 % respectively, while β-sheet increased to ~26 %. These changes indicate a temperature dependent transformation of BSA secondary structure, in particular the partial conversion of α-helix and β-turn components into β-sheets and potentially random coils [9], although the latter was not detected in our analysis. Our results are similar with previous reports of heat-induced denaturation of BSA. For example, Zhou et al. [10] observed that at temperatures above 65 °C there was a decrease in α-helix and extended chain structures, and an increase in the relative content of β-sheet structures. This agrees with our observation that secondary structure rearrangements occur only at elevated temperatures.

We did not detect any random coil structures at either temperature, which may be attributed to our assignment of the 1655 cm$^{-1}$ band to α-helix. This interpretation is in good agreement with previous reports where the 1654–1658 cm$^{-1}$ region is assigned to α-helices [4, 11], however, there have been inconsistencies in literature regarding the frequency assignment for α-helix. Some researchers attribute the frequency of 1660 cm$^{-1}$ to random coil structures [9, 12], whereas others suggest that α-helix and random coil overlap within the same absorption band frequency [9, 13], resulting in ambiguity in their distinction.

The band at 1627 cm$^{-1}$ was assigned based on literature [14], where H$_2$O was commonly used as the solvent. Although PBS was used in our study, after solvent subtraction, the spectra should not differ substantially [15]. Dong et al. [4] assigned bands in the region of 1624–1642 cm$^{-1}$ to β-sheets. However, the assignment of the band around 1630 cm$^{-1}$ has been under controversy in literature. While it is frequently attributed to intramolecular β-sheet structures [14, 16-18], several studies propose that it arises from short segment chains connecting α-helix segments [19, 20]. Given that the secondary structure of BSA consists of approximately 67% helix, 10% turn and 23% extended chain, and that x-ray crystallographic studies of HSA have revealed no β-sheet content [21-23], the interpretation of the 1630 cm$^{-1}$ band remains uncertain.

**SUPPLEMENTARY NOTE 2: QSUP Simulation**

The SPDC intensity of signal photons in our simulation is given by equation (3) in the main text with $\delta_s$ as

$$\delta_s = \delta_a + \delta_b + \delta_m \tag{S1}$$

with

$$\delta_a(\lambda_s, \theta_s) = \left(k_{p,a} - k_{s,a} \cos\theta_{s,a} - k_{i,a} \cos\theta_{i,a}\right) L_a, \tag{S2}$$

$$\delta_b(\lambda_s, \theta_s) = \left(k_{p,b} - k_{s,b} \cos\theta_{s,b} - k_{i,b} \cos\theta_{i,b}\right) L_b, \tag{S3}$$

$$\delta_m(\lambda_s, \theta_s) = \left(k_{p,m} - k_{s,m} \cos\theta_{s,m} - k_{i,m} \cos\theta_{i,m}\right) L_m. \tag{S4}$$

Here $\delta_m, \delta_b$ and $\delta_a$ are the longitudinal phase mismatch in the nonlinear crystal, including contributions from the sample, the biocell and air gap. The quantities $k_{\{p,s,i\},\{a,b,m\}}$ are the wavenumbers of the pump, signal and idler in the air gap, biocell, and material sample, with $\theta_{\{p,s,i\},\{a,b,m\}}$ denoting their corresponding propagation angles. Some parameters are fixed according to the specifications of the instruments we use. For example, $L_b$=10 mm is determined by the thickness of the CaF$_2$ windows in the biocell (the biosample holder) where each window has a thickness of 5 mm. However, several parameters influence the efficiency, fringe visibility, and fringe position on the camera/screen. The most critical parameters include the length of the nonlinear crystal, the sample optical path length, and the distance between the nonlinear crystal and the parabolic mirror (air gap in simulation). Understanding the role of each parameter is essential for optimizing experimental performance. Based on the simulated visibility, key experimental parameters are carefully chosen.

### A. The effect of nonlinear crystal length

The nonlinear crystal length L controls both the SPDC efficiency and the phase-matching bandwidth. As discussed in the main text, this introduces a trade-off between photon-pair generation efficiency and interference visibility.

Numerical simulations were performed for a range of crystal lengths, and L=5 mm was identified as an optimal value, providing a balance between photon flux and fringe visibility.

The visibility V obtained from the QSUP model and the transmissivity $\tau$ derived from FTIR are not intrinsically equal, as they originate from different measurement approaches. Their comparison is made under the assumption of equivalent sensitivity to absorption, such that variations in $\tau$ correspond to variations in V (see Supplementary Fig. S2).

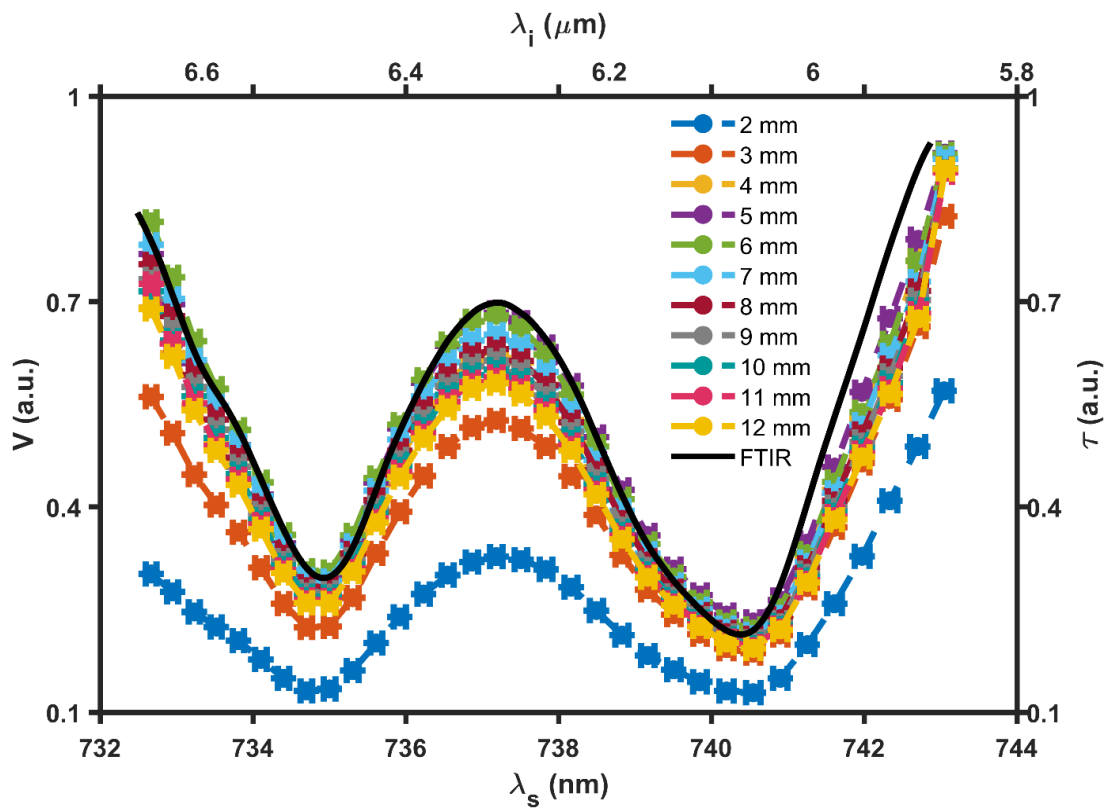

**Supplementary Figure S2**. The effect of different nonlinear-crystal lengths on visibility. L=5 mm is identified as optimal for this numerical experiment.

## B. The effect of sample optical path length

The sample length $L_m$ determines the optical path length and thus the interaction strength between the idler photon and the sample. As discussed in the main text, this leads to a trade-off between absorption and interferometric visibility.

Simulations indicate that the sample optical path length 6–8 μm provides near-optimal performance, with 6 μm selected to minimise sample volume. This corresponds to a sample volume of approximately 10 μL in a standard CaF2 biocell, which is smaller than that typically required for conventional FTIR measurements.

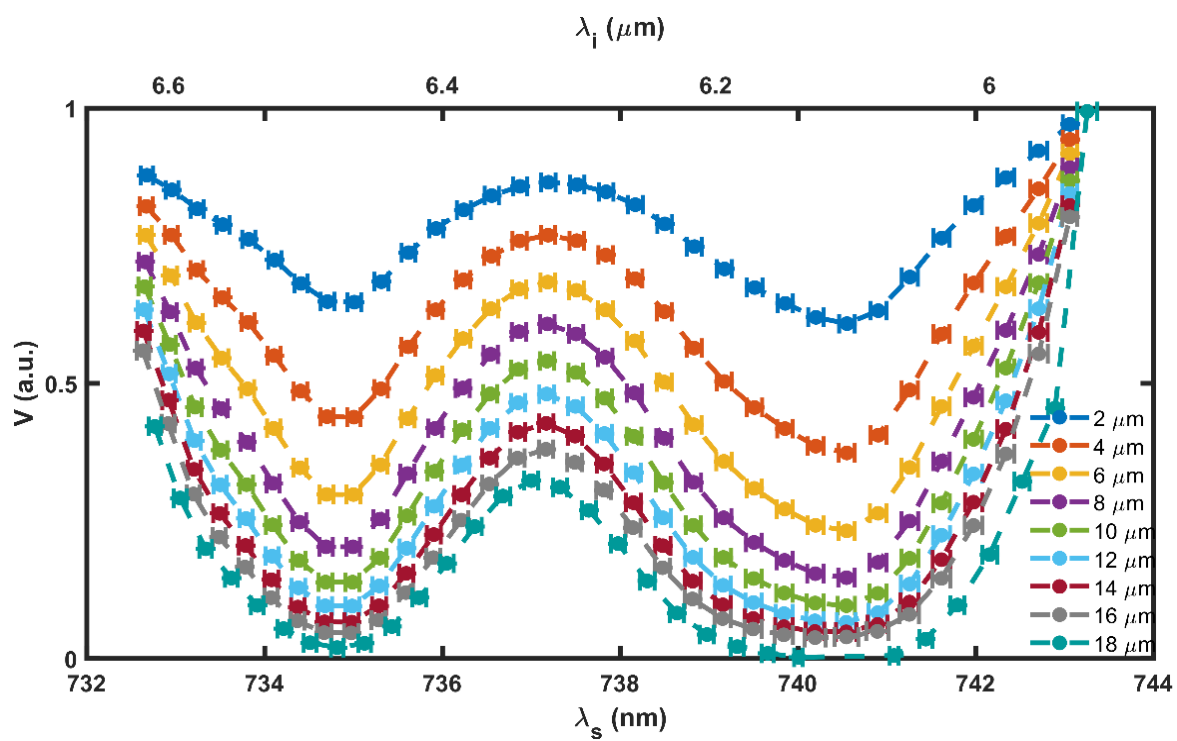

**Supplementary Figure S3**. The effect of different values of the sample lengths on visibility. $L_m$= 6–8 μm is identified as optimal for this numerical experiment.

## C. The effect of Crystal-to-OAPM Distance

The crystal–mirror separation $L_a$ controls the phase difference between the two SPDC passes and therefore the spectral position of the interference fringes (see main text). Simulations show that $L_a$=−8.75 mm provides optimal fringe alignment across the Amide I–II region, maximising spectral contrast.

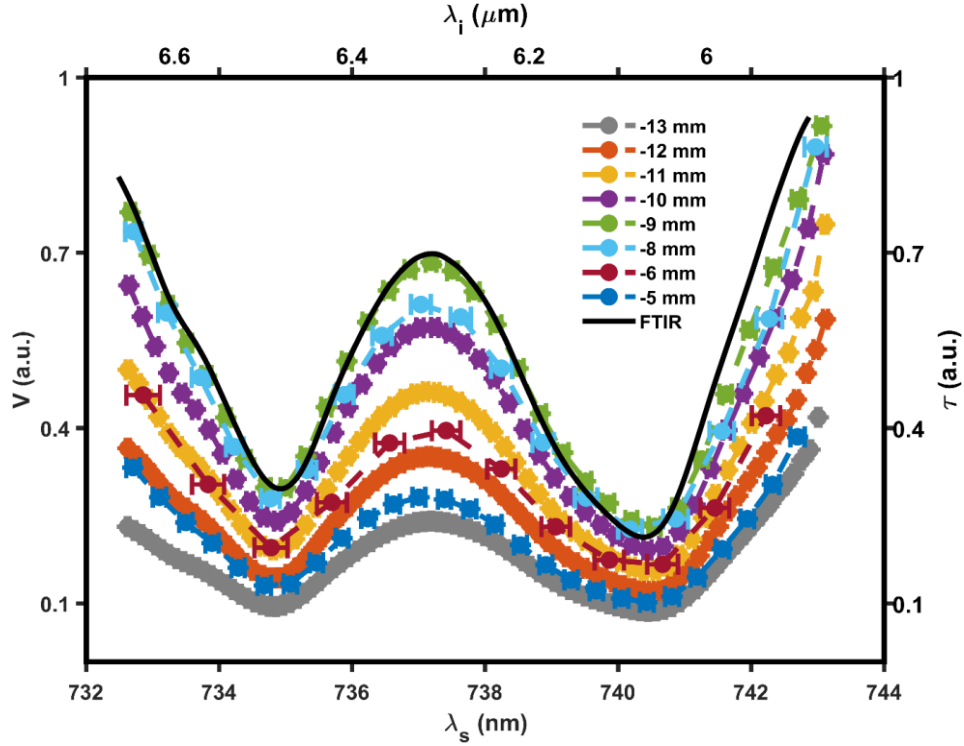

**Supplementary Figure S4**. The effect of various values of Crystal-to-OAPM Distance

However, the analysis also reveals that not all values of $L_a$ are suitable for reliable measurements. In particular, around $L_a \approx -7$mm, the interferometric visibility approaches zero. This behaviour can be observed in the fringe pattern as illustrated in Supplementary Fig. S5. For $L_a = -7$mm, shown in Supplementary Fig. S5 (a) the phase map of $\cos(\delta + \delta_a + \delta_b + \delta_m)$ shows strongly curved fringes that run almost parallel to the line given by phase matching in the crystal (black line in Supplementary Fig. S5 (a)). In contrast, for $L_a = -9$mm the phase fringes become nearly vertical, and the black line clearly crosses the fringes (see Supplementary Fig. S5 (b)).

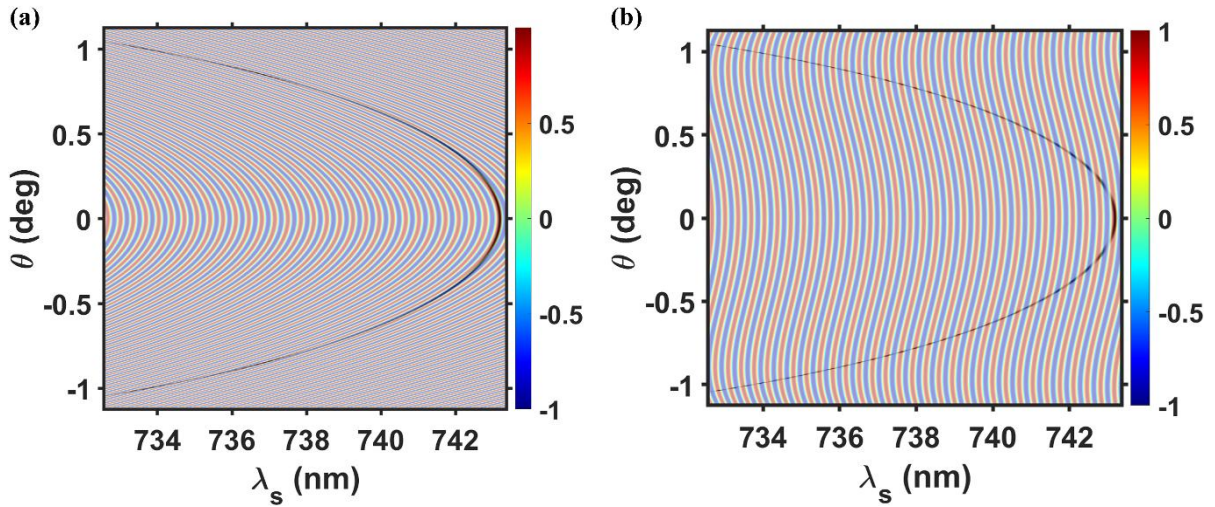

**Supplementary Figure S5**. Effect of crystal–mirror distance on interference fringes. Overlay of $\cos(\delta + \delta_a + \delta_b + \delta_m)$ with the angular–wavelength spectrum showing the change in fringe angular dependence. **(a)** $L_a = -7$ mm, where the visibility is suppressed. **(b)** $L_a = -8.75$ mm, where clear high-contrast fringes are observed.

The fringe visibility will shrink to exactly zero when the function describing the fringes $\cos(\delta + \delta_a + \delta_b + \delta_m)$ is constant along the line $\delta = 0$ in the $(\lambda_s, \theta_s)$ plane. This situation corresponds to the case where the phase accumulated along the various parts of the NL crystal, material, biocell and air gap cancel exactly along the optical path. An approximation for when this occurs can be obtained under the assumption that the transverse wavenumbers are small. For convenience, we define the transverse wavenumber as

$$q = k_s \sin \theta_s . \tag{S5}$$

From Equations (S2-S4), and noting that the transverse wavevector is conserved across each section, we have the following for the individual phases across the air gap, biocell and sample:

$$\delta_a = \left(k_{p,a} - \sqrt{k_{s,a}^2 - q^2} - \sqrt{k_{i,a}^2 - q^2}\right) L_a , \tag{S6}$$

$$\delta_b = \left(k_{p,b} - \sqrt{k_{s,b}^2 - q^2} - \sqrt{k_{i,b}^2 - q^2}\right) L_b , \tag{S7}$$

$$\delta_m = \left(k_{p,m} - \sqrt{k_{s,m}^2 - q^2} - \sqrt{k_{i,m}^2 - q^2}\right) L_m . \tag{S8}$$

The length of the sample material is typically much smaller than that of both the air gap and the biocell, being on the scale of microns as opposed to cm, i.e. $L_m \ll L_a, L_b$; for the purpose of examining the visibility of the fringes the phase accumulated across the sample can therefore be neglected. We furthermore now assume that all angles involved remain small on the phase-matching line $\delta = 0$, such that $q \ll k_s, k_i$ in both the air gap and the biocell. We can then expand the total phase change as:

$$\delta + \delta_a + \delta_b + \delta_m \approx \left[k_{p,a} - k_{s,a} - k_{i,a} + \frac{q^2}{2}\left(\frac{1}{k_{s,a}} + \frac{1}{k_{i,a}}\right)\right] L_a$$

$$+ \left[k_{p,b} - k_{s,b} - k_{i,a} + \frac{q^2}{2}\left(\frac{1}{k_{s,b}} + \frac{1}{k_{i,b}}\right)\right] L_b . \tag{S9}$$

Because of energy conservation in the SPDC process in the NL crystal we have

$$\frac{k_p}{n_p} - \frac{k_s}{n_s} - \frac{k_i}{n_i} = 0 , \tag{S10}$$

where $n_p$, $n_s$ and $n_i$ are the refractive indices of the NL crystal at the signal, pump and idler wavelengths respectively. Using this relation we find that $k_{p,a} - k_{s,a} - k_{i,a} = n_a \left(\frac{k_p}{n_p} - \frac{k_s}{n_s} - \frac{k_i}{n_i}\right) = 0$, and $k_{p,b} - k_{s,b} - k_{i,b} = n_a \left(\frac{k_p}{n_p} - \frac{k_s}{n_s} - \frac{k_i}{n_i}\right) = 0$ where $n_a$ is the refractive index of the gap and $n_b$ is the refractive index of the biocell, both assumed to be approximately constant across the entire wavelength range. We then have the total phase accumulation:

$$\delta + \delta_a + \delta_b + \delta_m \approx \frac{q^2}{2}\left(\frac{1}{k_{s,a}} + \frac{1}{k_{i,a}}\right)L_a + \frac{q^2}{2}\left(\frac{1}{k_{s,b}} + \frac{1}{k_{i,b}}\right)L_b$$

$$\approx \left(\frac{n_s}{k_s} + \frac{n_i}{k_i}\right)\left(\frac{L_a}{n_a} + \frac{L_b}{n_b}\right) \quad . \tag{S11}$$

The total phase is therefore constant, across all wavelengths and to fourth order in q, provided

$$\left(\frac{L_a}{n_a} + \frac{L_b}{n_b}\right) = 0 \tag{S12}$$

that is, for air gap lengths given by

$$L_a^{min} \approx -\frac{n_a}{n_b}L_b \tag{S13}$$

This is an approximate condition for the point near which the visibility of the fringes exactly vanishes over the entire wavelength range. The exact value of $L_a$ where the fringe visibility disappears may change if the angles $\theta_s$ and $\theta_i$ become larger, or if the sample length $L_m$ approaches the scale of the air gap and biocell.

### SUPPLEMENTARY REFERENCES